\documentclass[11pt,oneside,a4paper]{article}
\usepackage{graphicx}
\usepackage{amssymb}			
\usepackage{amsmath}			
\usepackage{amsfonts}
\usepackage{authblk}
\usepackage{subcaption}

\usepackage{algorithm}
\usepackage{algorithmic}

\begin{document}

\newtheorem{definition}{Definition}
\newtheorem{proposition}{Proposition}
\newtheorem{property}{Property}
\newtheorem{example}{Example}

\title{Finding overlapping communities\\in multiplex networks}
\author{Nazanin Afsarmanesh\\n.afsarmanesh@gmail.com\\Dept.~of Information Technology\\Uppsala University \and Matteo Magnani\\matteo.magnani@it.uu.se\\Dept.~of Information Technology\\Uppsala University}

\date{}

\maketitle

\begin{abstract}
We define an approach to identify overlapping communities in multiplex networks, extending the popular clique percolation method for simple graphs. The extension requires to rethink the basic concepts on which the clique percolation algorithm is based, including cliques and clique adjacency, to allow the presence of multiple types of edges.
\end{abstract}

\section{Introduction}

Community detection, also known as graph clustering, is one of the main tasks to study complex systems represented as networks.
A large number of community detection methods has appeared in the literature \cite{Fortunato2010,Coscia2011}, with early methods sharing two main features: they group the nodes of the network into a set of disjoint clusters --- also called \emph{partitioning clustering}, and they operate on \emph{simple graphs}, that is, graphs with at most one edge between each pair of nodes, no edges connecting a node to itself and no attributes (with the possible exception of weights on the edges).

However, simple graphs and partitioning clusterings do not accurately represent the complexity of several types of systems. For example, in social networks individuals communicate with different groups of people, like friends, colleagues, and family, and this determines multiple types of relationships between nodes, including multiple types of ties between the same pairs of nodes. In addition, people may belong to more than one community at the same time.

To increase the expressiveness of models based on simple graphs, multilayer networks \cite{Kivela2014,Dickison} and heterogeneous information networks \cite{Sun2012} have been introduced --- among other models, allowing nodes and edges to have different types and to be described by multiple attributes. A specific type of multilayer system, called \emph{multiplex network}, is characterized by nodes that can be connected through multiple types of edges and has been used for almost one century in the field of social network analysis \cite{Bott1928,Moreno1934}.
For what concerns community detection, people have developed several methods to find overlapping communities in simple graphs \cite{Xie2013}. Clique based methods \cite{palla2005uncovering,kumpula2008sequential,yan2009detecting}, fuzzy community detection algorithms  \cite{nepusz2008fuzzy,zhang2007identification} and link partitioning methods \cite{ahn2010link,evans2009line} are examples of overlapping clustering algorithms.

To the best of our knowledge, these two lines of research have not met yet: while we have methods for overlapping community detection on simple graphs \cite{Xie2013}, and we have partitioning community detection methods for multilayer networks \cite{Bothorel2015}, the problem of detecting overlapping communities in multilayer networks has not been directly addressed. Some approaches convert the multilayer network to a simple graph \cite{Berlingerio2011c,cai2005community,rodriguez2010exposing,tang2012community}, and then employ existing methods. However, this may result in information losses, because the clustering algorithm would not know whether a set of edges belongs to the same or to different layers, potentially leading to the discovery of communities scattered across a large number of layers and weak ties. Methods representing multiplex networks as multilayer structures, where the same node can belong to multiple layers, allow some type of natural overlapping because nodes in different layers are treated as separate entities and may end up in different clusters even when a partitioning algorithm is applied \cite{Mucha2010}. However, this type of overlapping does not allow a node inside a specific layer to belong to multiple clusters. This would prevent the identification of commonly encountered structures, like a person belonging to multiple working groups in a company, with a layer representing working relationships, or like the case of multiple interdependent social media networks modeled as multilayer networks \cite{DBLP:conf/asonam/MagnaniR11} where the different layers may contain several overlapping communities. In this paper we introduce an approach to identify overlapping communities in multiplex networks, where communities can both span multiple layers and overlap inside only one or more of them.

In the next section we present the basic definitions and concepts needed to understand our extended method: multiplex networks, partitioning and overlapping community structure, cliques and the original clique percolation method (CPM). Section~\ref{mCPM} extends the basic concepts on which the clique percolation algorithm is based, including cliques and clique adjacency, to allow the presence of multiple types of edges. In this section we highlight how well-understood concepts like cliques can be extended in different ways when multiple layers are considered. In Section~\ref{algorithm} we go through the main algorithmic steps used to compute communities, that we also exemplify on a small network, and in Section \ref{experiments} we present an experimental characterization of the communities that can be identified using our method.

\section{Preliminaries}

\subsection{Multilayer and multiplex networks}

\begin{figure}[ht]
\centering
\begin{subfigure}{.3\textwidth}
    \includegraphics[width=\textwidth]{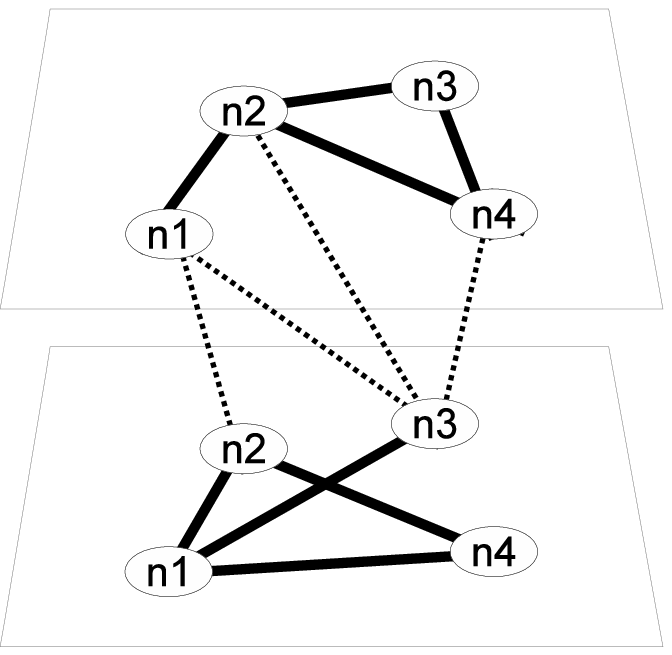}
    \caption{A multilayer network}
    \label{fig:ml1}
\end{subfigure}
\begin{subfigure}{.3\textwidth}
    \includegraphics[width=\textwidth]{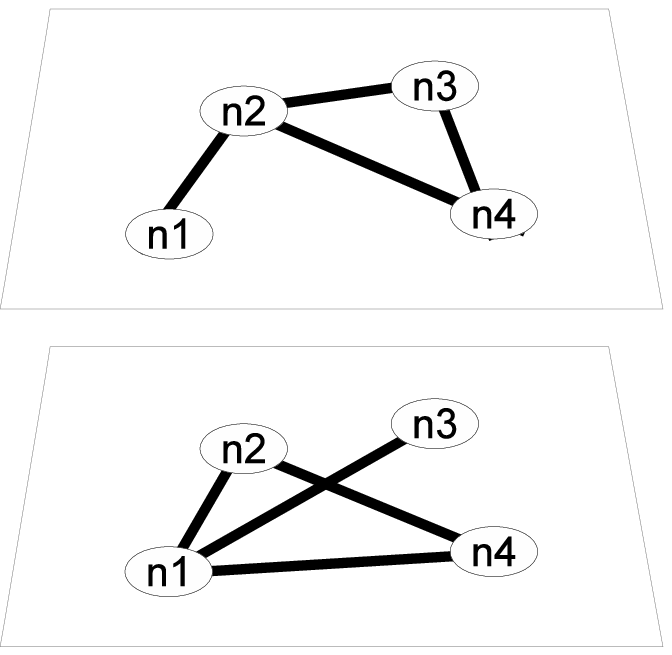}
    \caption{A multiplex network}
    \label{fig:ml2}
\end{subfigure}
\begin{subfigure}{.29\textwidth}
    \includegraphics[width=\textwidth]{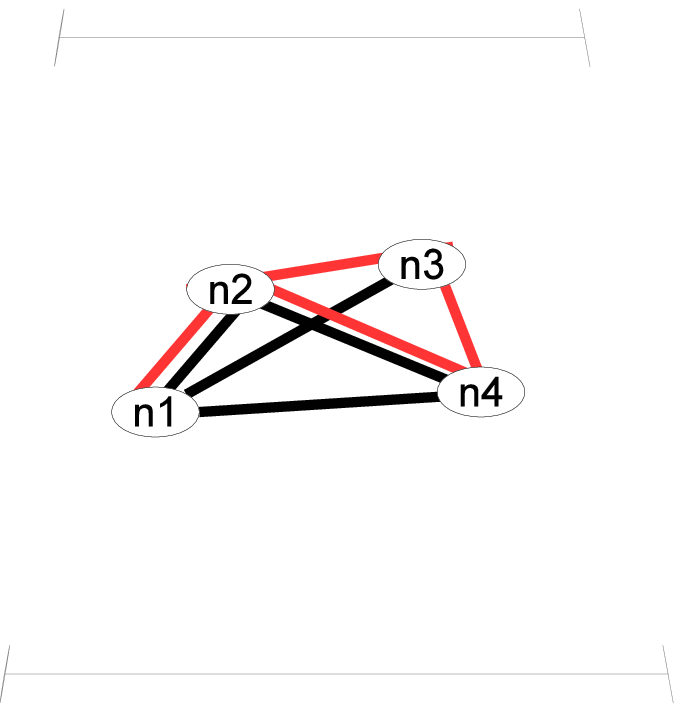}
    \caption{A multi-graph}
    \label{fig:ml3}
\end{subfigure}
\caption{Network models}
\label{fig:ml}
\end{figure}

Multilayer networks are data structures where the same node can belong to multiple contexts called layers. An example with four nodes and two layers is shown in Figure~\ref{fig:ml1}.

\begin{definition}[Multilayer network]\index{multilayer network model}
Given a set of nodes $\mathcal{N}$ and a set of layers $\mathcal{L}$, a multilayer network is defined as a quadruple $M = (\mathcal{N},\mathcal{L},V,E)$ where $(V,E)$ is a graph and $V \subseteq \mathcal{N} \times \mathcal{L}$.
\end{definition}

In this paper we focus on a specific type of multilayer network called multiplex network, where edges can only exist between nodes in the same layer. We can represent multiplex networks separating the nodes into different layers, as in Figure~\ref{fig:ml2}, or use an alternative representation as an edge-labeled multi-graph, as in Figure~\ref{fig:ml3}, where different colors represent the different layers.

\begin{definition}[Multiplex network]\index{multilayer network model}
A multiplex network is a multilayer network where $((n_1,l_1),(n_2,l_2)) \in E$ implies that $l_1 = l_2$.
\end{definition}

\subsection{Communities in multilayer networks}

Given a multiplex network $M = (\mathcal{N},\mathcal{L},V,E)$, we can group its nodes $\mathcal{N}$ into $q$ sets $C = \{C_1, \dots, C_q\}$, where we allow different groups to overlap. For example, in Figure~\ref{fig:ml2} we can group the nodes in two sets $C_1 = \{n1, n2, n4\}$ and $C_2 = \{n2, n3, n4\}$. Among the many possible ways of grouping nodes, we want to find one  representing the \emph{(overlapping) community structure} of the network.

Some methods, like \cite{newman2004finding}, use a quality function to compare different ways of assigning the nodes to groups, for example assigning a higher quality to solutions where nodes that are connected together are included in the same group. In other cases, the shape of the community structure is defined by the specific method used to discover it. For example, the clique percolation method described in the next section provides a specific definition of overlapping community structure based on the concept of clique.

\subsection{Clique percolation}

The clique percolation method
was introduced by Palla et al.~in 2005 \cite{palla2005uncovering}. For a given $k$, CPM builds up communities from $k$-cliques, that is, complete subgraphs in the network with $k$ nodes. Two $k$-cliques are said to be adjacent if they share $k-1$ nodes. A \textit{$k$-clique community} is defined as a maximal union of $k$-cliques that can be reached from each other through a series of adjacent $k$-cliques. In general, if the number of links is increased above some critical point, a giant community would appear that covers a vast part of the system. Therefore, $k$ is chosen as the smallest value where no giant community appears. CPM allows overlapping communities in a natural way as a node can belong to multiple cliques.

\begin{figure}
\centering
\includegraphics[width=.6\textwidth]{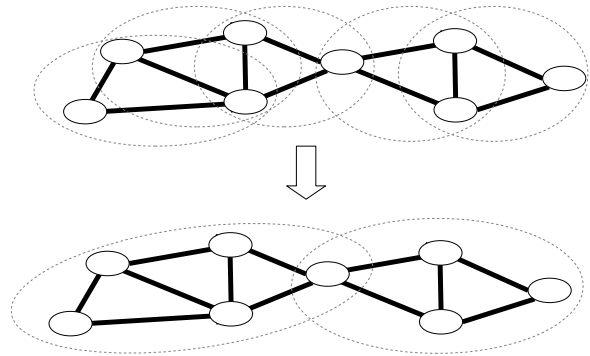}
\caption{A schematic view of the CPM method}
\label{fig:CPM}
\end{figure}

Figure~\ref{fig:CPM} shows a simplified example of how CPM works. Given a simple graph, first cliques are identified (in this example, we only have 3-cliques for simplicity), then adjacent cliques are grouped together to form two communities with one common node.

\section{Multiplex clique percolation} \label{mCPM}

Our extended CPM algorithm for multiplex networks (CPM$^M$), of which we describe an implementation in the next section, follows the same main general steps of CPM. However, the concepts on which it is based must be extended to multiplex networks. In particular, we need to define:
\begin{enumerate} 
\item What a clique on multiple layers/edge types is.
\item When two multiplex cliques can be considered adjacent.
\item How adjacent cliques should be grouped to build communities.
\end{enumerate}

\subsection{Cliques on multiple layers}

\begin{figure}
\centering
\includegraphics[width=0.7\textwidth]{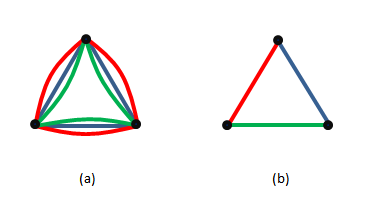}
\caption{Multiplex cliques}
\label{fig:Cliques}
\end{figure}

While a clique on a simple graph is a well understood structure, defined as a set of nodes that are all connected to each other, the same concept can be extended in different ways for multiplex networks depending on how multiple edge types can contribute to the clique connectivity. Considering a specific number of layers, we might require that a clique contains all the possible edges on all these layers. In other words, a clique is formed by a combination of cliques in individual layers. We refer to this type of cliques as AND-cliques. On the other hand, we might consider it sufficient if each pair of nodes is connected on at least one layer to form a clique in the multilayer network. This type of cliques can be referred to as OR-cliques. The graph in Figure~\ref{fig:Cliques}(a) is an AND-clique on 3 layers, while the one in Figure~\ref{fig:Cliques}(b) is an OR-clique on the same layers. In this paper, we focus on AND-cliques.

\begin{definition}[k-m-AND-clique]
Let $L_{ij}$ be the set of edge types (layers) between nodes $i$ and $j$. We define a k-m-AND-clique as a subgraph in the multilayer network with $k$ nodes that includes a combination of at least $m$ different $k$-cliques coming from $m$ different layers. In other words, a k-m-AND-clique is a subgraph with $k$ nodes $C$ where 
\begin{equation}\label{CIR}
\small |\bigcap\limits_{i,j\in C} L_{ij}| \geq m
\end{equation}
\end{definition}

Similar to the case of cliques on simple graphs, we can have the concept of maximality of cliques in multilayer networks. An induced subgraph $C$ is a maximal $k$-$m$-AND-clique if: 1) $C$ is a $k$-$m$-AND-clique, 2) There is no $m'>m$ such that $C$ is also $k$-$m'$-AND-clique, or in other words, $C$ is maximal on $m$, and 3) $C$ is not included in any $k'$-$m$-AND-clique where $k'>k$, in other words, $C$ is maximal on $k$.

\subsection{Adjacency and communities}

\begin{figure}
\centering
\includegraphics[scale=0.9]{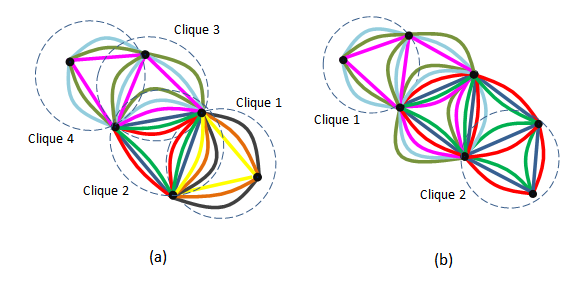}
\caption{Adjacent cliques}
\label{fig:adjacency}
\end{figure}

When cliques may exist on different layers, the concept of adjacency should also consider this aspect. 
To illustrate why, consider a definition of adjacency where $k$-$m$-AND-cliques only need to share $k-1$ nodes to be considered adjacent. Figure~\ref{fig:adjacency}(a) shows a series of adjacent 3-3-AND-cliques. As we see, adjacent cliques do not necessarily share any edge types on all pairs and they might share edge types only on their common pairs of nodes (e.g., cliques 1 and 2). It is worth noting that more diversity among the edge types in external connections of adjacent cliques results in denser internal connectivity. In addition, cliques at distance one still have to share some edge types on some of their pairs of nodes (e.g., cliques 1 and 3 in Figure~\ref{fig:adjacency}(a)), however, when the distance between cliques becomes greater than one, they might have completely different edge labels (cliques 1 and 4 in Figure~\ref{fig:adjacency}(a)).

To guarantee more uniformity among edge types we should then introduce an additional constraint on the edge labels for clique adjacency. Two $k$-$m$-AND-cliques are said to be $m'$-adjacent if they share $k-1$ nodes and they also share at least $m'$ edge types on all of their pairs of nodes.

However, interestingly this constraint is not sufficient in itself to guarantee uniformity at community level, that is, on a maximal set of adjacent cliques. Figure~\ref{fig:adjacency}(b) shows a series of 3-adjacent 3-3-AND-cliques. As we see, although this constraint enforces uniformity among edge labels in small neighborhoods, it cannot guarantee the presence of common edge types for all cliques which are reachable from each other, e.g., cliques 1 and 2 in Figure~\ref{fig:adjacency}(b). While this example appears to be quite complex and has been specifically constructed to highlight this unwanted behavior, it nevertheless shows that this situation is theoretically possible and should be taken care of.

To enforce uniformity among edge types throughout the whole community, we need more constraints than constraints on cliques' adjacency. Then, we define a [$(k-m)$-AND-clique]$_{(m',m'')}$ community as the maximal union of $m'$-adjacent $k$-$m$-AND-cliques where all cliques share at least $m''$ edge types on all of their pairs of nodes. Therefore, a [$(k-m)$-AND-clique]$_{(m',m'')}$ community is a group of nodes where the nodes form $k$-clique communities on at least $m''$ different layers. Please notice that this is a very general definition, and in practice we can just use two parameters: $k$ and $m (= m, m', m'')$. However, in future work we will explore whether it can be practically valuable to identify communities sharing only some of the layers constituting their cliques or determining adjacency.

\section{Algorithm}\label{algorithm}

\begin{figure}[ht]
\centering
\begin{subfigure}{.2\textwidth}
    \includegraphics[width=\textwidth]{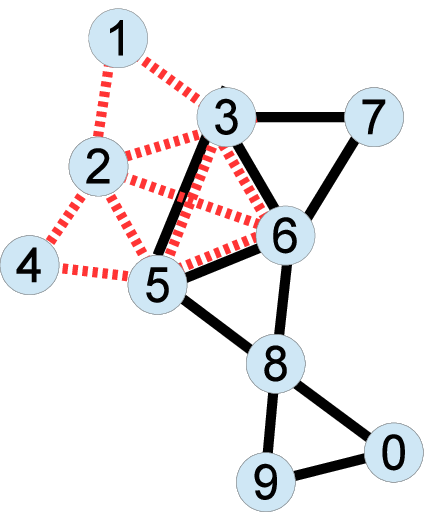}
    \caption{Input graph}
    \label{fig:example1}
\end{subfigure}
\begin{subfigure}{.5\textwidth}
    \includegraphics[width=\textwidth]{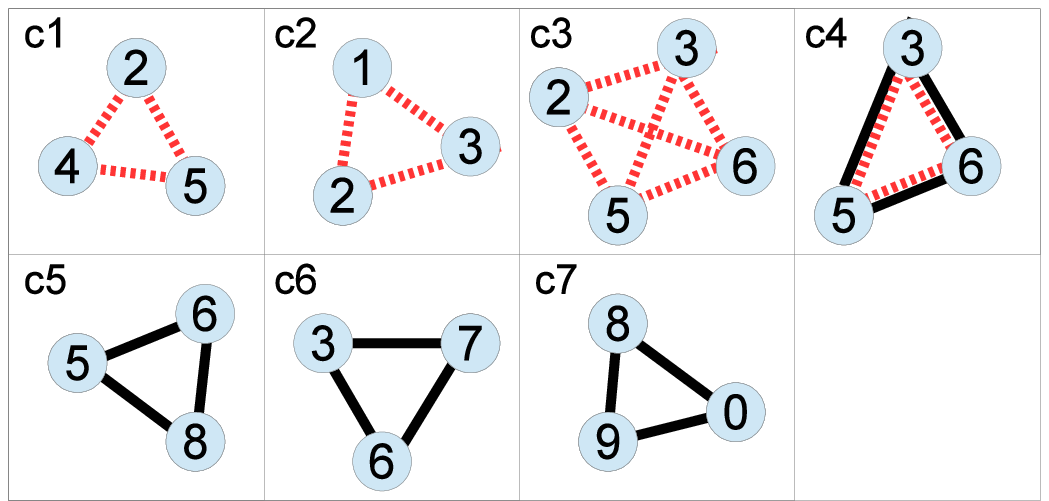}
    \caption{Maximal cliques}
    \label{fig:example2}
\end{subfigure}
\begin{subfigure}{.2\textwidth}
    \includegraphics[width=\textwidth]{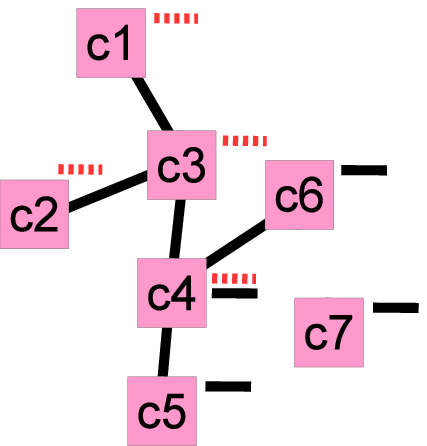}
    \caption{Adjacency}
    \label{fig:example3}
\end{subfigure}
\begin{subfigure}{.45\textwidth}
    \includegraphics[width=\textwidth]{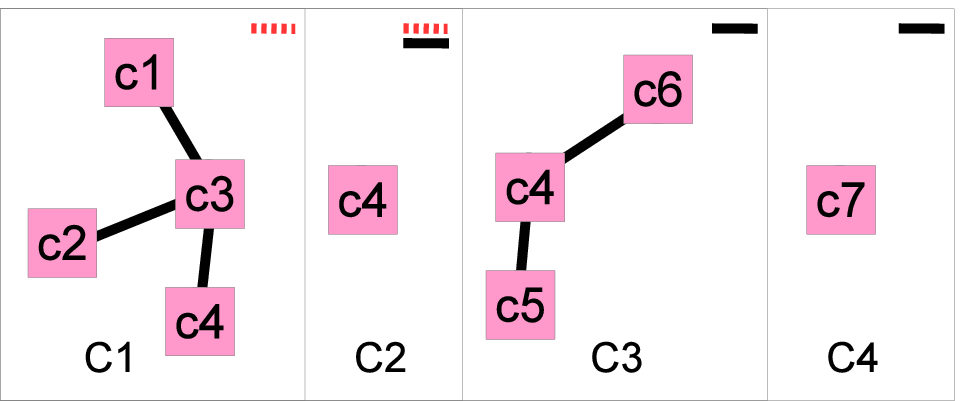}
    \caption{Maximal adjacent structures}
    \label{fig:example4}
\end{subfigure}
\begin{subfigure}{.5\textwidth}
    \includegraphics[width=\textwidth]{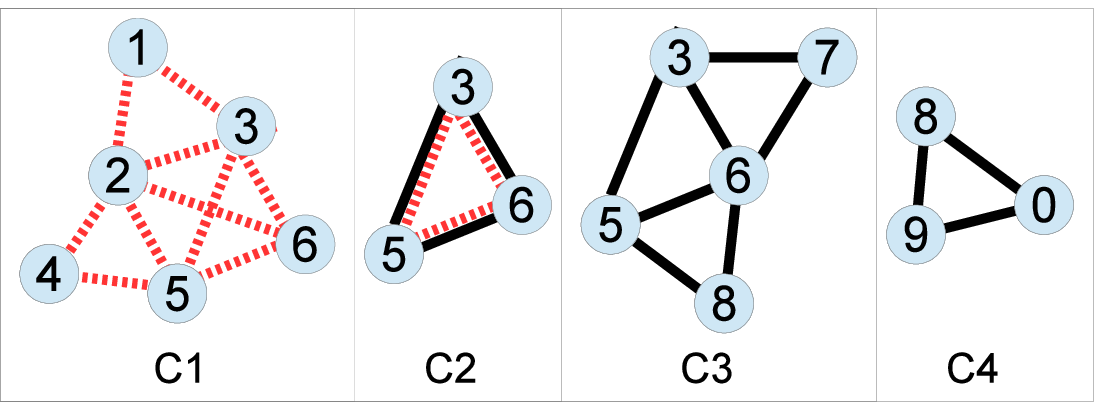}
    \caption{Corresponding clusters}
    \label{fig:example5}
\end{subfigure}
\caption{A step-by-step view of our approach}
\label{fig:example}
\end{figure}

In this section we present an algorithm to detect communities according to our definitions. We will focus on the strongest definition of community, based on AND cliques, and without loss of generality we will assume that $m=m'=m''$. The algorithm is divided into three parts, as in the original method: finding cliques, building the adjacency graph and extracting communities --- notice that in the original method an overlap matrix is constructed in the intermediate step, but the role of this structure is analogous to the role of our adjacency graph. These three parts are described in the following, and exemplified in Figure~\ref{fig:example}.

\subsection{Locating the cliques}
Our method is based on first locating all maximal $k$-$m$-AND-cliques. As an example, in Figure~\ref{fig:example2} we show all maximal cliques with $k \geq 3$ and $m \geq 1$ extracted from the network in Figure~\ref{fig:example1}. Please notice that for the cliques in Figure~\ref{fig:example2} we have only represented the $m$ edge types where the clique is defined, that is, the $m$ edge types that are present between all pairs of nodes in the clique.

Algorithm~\ref{al:find-cliques} is an extension of Bron--Kerbosch's algorithm designed to perform this step. It is a recursive algorithm where the recursion step takes a clique $A$ as input and returns all maximal $k$-$m$-AND-cliques containing $A$ that can be constructed using nodes in $B$, with $k \geq \overline{k}$ and $m \geq \overline{m}$. In this way, given a multiplex network $M = (\mathcal{N},\mathcal{L},V,E)$, a call to  
$\textrm{find-cliques}(\{\}, \mathcal{N}, \{\}, \overline{k}, \overline{m})$ with $\overline{k}>1$ and $\overline{m}>0$ returns all maximal cliques in $M$ with $k \geq \overline{k}$ and $m \geq \overline{m}$.

The algorithm works by updating two sets: one containing nodes that can be used to extend the currently processed clique, and one to keep track of already examined cliques. More precisely, the parameter $B$ is a set of nodes such that, for every node $n \in B$, $A \cup \{n\}$ is a previously unseen clique on at least $\overline{m}$ layers. Whenever a node from $B$ is used to extend the clique $A$, then $B$ is updated by removing those nodes that are no longer connected to all nodes in the new clique $A'$. $C$ is the same as $B$, but containing those nodes that have already been examined by the algorithm during some previous iteration, so that no duplicates are produced. Given a set of nodes $A$ we notate the set of layers where the nodes in $A$ form a clique as $L(A)$, and the number of nodes in $A$ as $S(A)$. Therefore, if $|L(A)|=0$ then $A$ is not a clique on any layer. We also define $\max(\emptyset) = 0$.

As an example, assume we call $\textrm{find-cliques}(\{\}, \{0, 1, \dots, 9\}, \{\}, 3, 1)$ on the network in Figure~\ref{fig:example1}. The algorithm would then start exploring one of the nodes in B, let us say 5. The new call will thus include in $B$ only those nodes that can still form a clique on at least one layer when joined with 5: $\textrm{find-cliques}(\{5\}, \{2, 3, 4, 6, 8\}, \{\}, 3, 1)$. Let us assume that at the next iteration 3 is added to the current clique: $\textrm{find-cliques}(\{3, 5\}, \{2, 6\}, \{\}, 3, 1)$, and then 6: $\textrm{find-cliques}(\{3, 5, 6\}, \{2\}, \{\}, 3, 1)$. At this point $S(A)$ is 3, satisfying the minimum clique size, and $|L(\{3, 5, 6\})| = 2 > \max(\{|L(A \cup \{b\})| : b \in B\}) = \max(\{|L(\{2, 3, 5, 6\})|\}) = 1$. In fact, $\{3, 5, 6\}$ is a clique on two layers, while $\{2, 3, 5, 6\}$ is a clique on only one layer. Therefore, the current clique is returned as a maximal one (c4 in Figure~\ref{fig:example2}). At the next iteration, $\textrm{find-cliques}(\{2, 3, 5, 6\}, \{\}, \{\}, 3, 1)$ is called and clique c3 is returned. At some later point, the algorithm would call $\textrm{find-cliques}(\{2, 3, 5\}, \{\}, \{6\}, 3, 1)$, not returning any clique because $C = \{6\}$ indicates that this path has already been explored, and $\textrm{find-cliques}(\{4, 5\}, \{2\}, \{3\}, 3, 1)$, ultimately leading to the discovery of clique c1, and so on.

\begin{algorithm}
\begin{algorithmic}[1]
\IF{$S(A) \geq \overline{k} \land \max(\{|L(A \cup \{b\})| : b \in B\}) < |L(A)| \land \max(\{|L(A \cup \{c\})| : c \in C\}) < |L(A)|$}
	\STATE{OUTPUT A}
\ENDIF
\FOR{$b \in B$}
	\STATE $A' = A \cup \{b\}$
	\STATE $B = B \setminus \{b\}$
	\STATE $B' = \{ b' \in B : |L(A' \cup \{b'\})| \geq \overline{m} \}$
	\STATE $C' = \{ c' \in C : |L(A' \cup \{c'\})| \geq \overline{m} \}$
	\STATE $\textrm{find-cliques}(A', B', C', \overline{k}, \overline{m})$
	\STATE $C = C \cup \{b\}$
\ENDFOR
\end{algorithmic}
\caption{$\textrm{find-cliques}(A, B, C, \overline{k}, \overline{m})$
}
\label{al:find-cliques}
\end{algorithm}

\subsection{Clique-adjacency graph}
In a simple graph, each clique is included in exactly one community, therefore, communities can be identified from a clique-clique overlap matrix (see \cite{palla2005uncovering} for the details). However, this statement is not necessarily true for $k$-$m$-AND-cliques and the corresponding communities.
Because of the more complicated relations between cliques, instead of the overlap matrix used in the original method we generate a related but different data structure.

This step is the simplest in the algorithm: in our running example, it produces the adjacency graph represented in Figure~\ref{fig:example3}, where each node of the graph corresponds to a maximal clique and an edge between two nodes indicates that the corresponding cliques share at least $k$ nodes and at least $m$ edge types on all of their pairs of nodes. In the graph we have indicated for each node the layers where the corresponding clique is defined. In the following we refer to this graph as clique-adjacency graph.




\subsection{From the adjacency graph to communities}
As previously mentioned, each clique can be included in different communities with different combinations of its adjacent cliques. Here our objective is using the adjacency graph and the information regarding the edge labels simultaneously to find communities in the multilayer network. 
Two cliques can be included in at least one community if: 1) there exists a path between the corresponding nodes in the graph, and 2) for all nodes in the path the corresponding cliques share at least $m$ edge types on all of their pairs. We call the latter rule the \textit{cliques' constraint}, and we call a tree \emph{maximal} if no other adjacent clique can be added without reducing the maximal $m$ for which the cliques' constraint is satisfied.
Therefore, each community in the original network corresponds to a maximal tree in the clique-adjacency graph where the cliques' constraint holds for all nodes in the tree. So the problem is equivalent to recognizing all such maximal trees in the graph.

Figure~\ref{fig:example4} shows all the maximal trees from our clique-adjacency graph for $m \geq 1$. 
As we see, clique $c4$ can be included in three communities: C1, C2 and C3. No new clique can be added to these sets without reducing the value of $m$ for which the cliques' constraint holds. As an example, community C4 satisfies the cliques' constraint for $m=2$. Adding any adjacent clique to it, like c3, c5 and c6, the constraint would no longer hold for $m=2$ because only one layer would be common for both cliques. In Figure~\ref{fig:example4} for each maximal tree we have indicated the layers where the constraint is satisfied, and Figure~\ref{fig:example5} shows all communities in this example for $m \geq 1$. 

Algorithm~\ref{al:find-communities} takes a community $A$ as input and returns all maximal communities containing $A$ with at least $m$ common layers, for $m \geq \overline{m}$. More precisely, the input $B$ is the set of cliques $c$ such that $A \cup \{c\}$ is a previously unseen community on at least $\overline{m}$ layers, while $C$ is the set of cliques $c$ such that $A \cup \{c\}$ is an already processed community on at least $\overline{m}$ layers, that is, a community that has already been examined by the algorithm during some previous iteration. The role of $B$ and $C$ is the same as in the previous algorithm, with the difference that in Algorithm~\ref{al:find-cliques} $B$ and $C$ can shrink at each recursive step, while in this algorithm they can grow because adding a node to the tree makes all the neighbors of this node candidate extensions. To decide whether a new clique should be included into $B$ or $C$, the algorithm keeps track of previously processed cliques ($D$). The difference between $C$ and $D$ is that $C$ includes only those cliques in $D$ that can be used to extend the current community. Given a set of cliques (that is, a community) $A$ we notate the set of layers common to all cliques in $A$ as $L(A)$. Therefore, if $|L(A)|=0$ then there is no layer common to all cliques. We also define $N(c)$ as the neighbors of $c$ in the adjacency graph.
Algorithm~\ref{al:find-communities-all} iterates over all cliques and finds all maximal communities containing the clique under examination for any $m \geq \overline{m}$ --- therefore, at the end it outputs all maximal communities.

\begin{algorithm}
\begin{algorithmic}[1]
\STATE $D = \emptyset$
\FOR{$c \in \textrm{Cliques}$}
	\STATE $\textrm{find-communities}(\{\}, \{c\}, \{\}, D, \overline{m})$
	\STATE $D = D \cup \{c\}$
\ENDFOR
\end{algorithmic}
\caption{$\textrm{find-communities}(\textrm{Cliques}, \overline{m})$
}
\label{al:find-communities-all}
\end{algorithm}

\begin{algorithm}
\begin{algorithmic}[1]
\IF{$S(A) > 0 \land \max(\{|L(A \cup \{b\})| : b \in B\}) < |L(A)| \land \max(\{|L(A \cup \{c\})| : c \in C\}) < |L(A)|$}
	\STATE{OUTPUT A}
\ENDIF
\FOR{$b \in B$}
	\STATE $A' = A \cup \{b\}$
	\STATE $B' = \{ b' \in B : b' \neq b \land |L(A') \cap L(\{b'\})| \geq \overline{m} \land b' \notin D \}$
	\STATE $C' = \{ c' \in B : c' \neq b \land |L(A') \cap L(\{c'\})| \geq \overline{m} \land c' \in D \}$
	\STATE $B' = B' \cup \{ b' \in N(b) : b' \neq b \land b' \notin A' \land |L(A') \cap L(\{b'\})| \geq \overline{m} \land b' \notin D \}$
	\STATE $C' = C' \cup \{ c' \in N(b) : c' \neq b \land c' \notin A' \land |L(A') \cap L(\{c'\})| \geq \overline{m} \land c' \in D \}$
	\STATE $C' = C' \cup \{ c' \in C : |L(A') \cap L(\{c'\})| \geq \overline{m} \}$
	\STATE $\textrm{find-communities}(A', B', C', D, \overline{m})$
	\STATE $D = D \cup \{b\}$
\ENDFOR
\end{algorithmic}
\caption{$\textrm{find-communities}(A, B, C, D, \overline{m})$
}
\label{al:find-communities}
\end{algorithm}

\section{Experiments}\label{experiments}

The objective of this section is to characterize the types of communities that our method is designed to identify.

\begin{figure}
\centering
\includegraphics[scale=0.7]{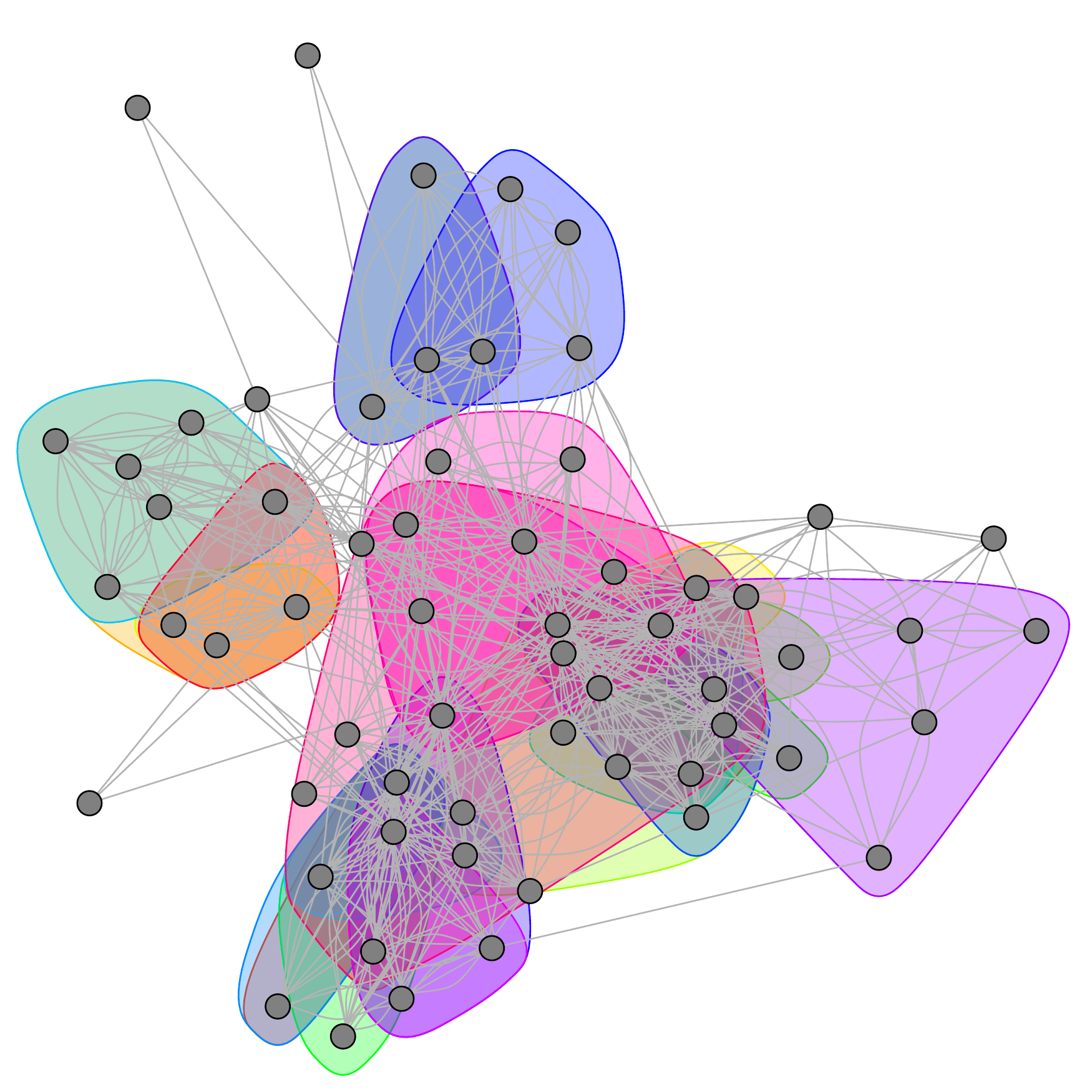}
\caption{Clusters in a real network}
\label{fig:clusters}
\end{figure}

Table 1 shows the result of our experiments on a real multilayer network of five layers \cite{Rossi2015} with k=3 and m=2 --- therefore, we only report communities that share at least two edge types, not those that can be found in the single layers one by one. As we see, we have found 26 communities where the size of communities vary from 3 to 12 nodes.  The layers Facebook, Work and Lunch which are denser than Leisure and Coauthor appear more frequently among the communities.  In Figure~\ref{fig:clusters} we can see how a few of the 61 nodes in the network are not included in any community.

It can be realized from this table that we can identify different types of overlapping nodes among communities in multilayer networks. If we consider the communities that are forming on the two layers Lunch and Work, e.g. C3 and C5, the structure of communities are more or less similar to the case of single networks as the communities are well-separated with a limited number of overlapping nodes. This can be considered as a general rule for the case where the sets of contributing layers are the same. On the other hand, we have two communities C21 and C22 where the sets of contributing layers are not exactly the same. These two communities have 5 overlapping nodes while C21 has only 6 nodes. This is in fact a consequence of what we experienced earlier, that is, cliques can be included in different communities forming by different combinations of layers. In addition, we can also have hierarchical community structures: small communities with larger number of contributing layers within larger communities with a smaller number of contributing layers, like C16 and C17.

\begin{table}
\begin{tabular}{l l l}
\hline
\# & nodes & layers \\
\hline
C01 & U107 U1 U29 U32 & facebook lunch \\
C02 & U124 U109 U47 & facebook lunch \\
C03 & U130 U134 U4 & lunch work \\
C04 & U91 U65 U72 & lunch work leisure \\
C05 & U4 U112 U68 U141 & lunch work \\
C06 & U59 U91 U110 U113 & facebook work leisure \\
C07 & U91 U65 U67 U72 & lunch work \\
C08 & U18 U62 U76 & lunch leisure \\
C09 & U107 U29 U32 & facebook lunch work \\
C10 & U22 U26 U42 U49 & lunch work \\
C11 & U107 U1 U29 U32 U17 U14 \textbf{+ 3} & lunch work \\ 
C12 & U91 U110 U53 & coauthor lunch work \\
C13 & U59 U91 U110 U53 & lunch work \\
C14 & U106 U118 U26 U41 & lunch work \\
C15 & U109 U54 U76 U79 & facebook lunch leisure \\
C16 & U59 U91 U110 & facebook lunch work leisure \\
C17 & U59 U91 U110 U113 U138 & work leisure \\
C18 & U1 U17 U14 U19 U23 U73 & lunch work leisure \\
C19 & U33 U123 U97 U71 U4 U67 U63 & lunch work \\
C20 & U123 U71 U4 U67 & facebook lunch work \\
C21 & U109 U18 U3 U54 U76 U79 & facebook lunch \\
C22 & U109 U126 U3 U54 U76 U79 U90 & lunch leisure \\
C23 & U106 U118 U41 & lunch work leisure \\
C24 & U123 U59 U71 U91 U130 U47 \textbf{+ 6} & facebook work \\ 
C25 & U130 U47 U99 & coauthor work \\
C26 & U59 U91 U126 U110 & lunch leisure \\
\hline
\end{tabular}
\caption{Communities identified in a real data set for k=3 and m=2}
\label{tab:aucs}
\end{table}

These overlapping structures identified by our method, with core communities built on many layers and larger communities including more peripheral nodes on less layers, are compatible with the type of communities often observed in online social networks, as studied by \cite{Leskovec2008}. Figure~\ref{fig:clusters} shows some of these structures with central dense areas corresponding to multiple overlapping communities on many layers and external whiskers containing nodes only contained inside communities on less layers. 

\section{Discussion}

In this paper we have extended the CPM method to identify overlapping communities in multiplex networks. We have first focused on the formal definition of the method, discussing how to extend existing concepts to the multiplex context, defined an algorithm and studies its empirical behavior on a real dataset.

Despite the need for a more thorough evaluation, some interesting aspects already emerge from the formal definition of the method and from our preliminary experiments.

\begin{itemize}
\item Multiple extensions can be defined based on how we want the different layers to contribute to the community structure. Our two extensions of the concept of clique (AND and OR) are related to each other, and compatible with the literature on multilayer networks that has already considered these two alternative interpretations of the interactions between different layers. 
\item Clique adjacency must consider both the nodes and the edge types.
\item Constraining clique adjacency to a maximum number of different edge types, so that different cliques contain some common edge types, is not sufficient to enforce uniformity at the community level, that is, inside the same group of adjacency-reachable cliques there can be cliques not sharing any edge type. To obtain more homogeneous communities we then need to define a limit to the heterogeneity we want to accept. 
\item The attempt to keep communities homogeneous results in a phenomenon not visible when single graphs and the original method are used. While in CPM the same node can belong to multiple communities, in CPM$^M$ whole cliques can belong to different communities, as it has been practically demonstrated in our working example. In CPM, either only clique c4 or the whole connected component of the adjacency graph containing c4 would be returned, resulting in a single big community not further separated into two overlapping ones. This shows an example where not considering the information about the different edge types would result in the detection of larger communities scattered on several layers, and also the fact that the type of overlapping produced using our approach enables the identification of some kind of hierarchical community structures, to be further investigated.
\end{itemize}

We are currently in the process of testing our algorithm on a wide range of datasets, to characterize its computational complexity and improving our understanding of the quality and kind of clusters it can identify. From the formalization of the method, we can see that it is at least as complex as CPM in the worst case. Our hypothesis, that we are currently testing, is that CPM$^M$ has a similar practical behavior, where the sparsity of the networks and their community structures make the practical execution time acceptable in real cases when either the size of the network or the minimum number of layers $m$ are not too large.

\bibliographystyle{unsrt}
\bibliography{bibliography,msna,sna,mine}

\end{document}